\documentstyle[eqsecnum,pre,preprint,aps,epsfig]{revtex}

\tightenlines

\newcommand{\boa}{{\bf a}}

\newcommand{\ens}{{\cal E}}

\newcommand{\be}{\begin{equation}}

\newcommand{\ee}{\end{equation}}
\newcommand{\bs}{\begin{mathletters}} 
\newcommand{\es}{\end{mathletters}} 

\newcommand{\baa}{\begin{eqnarray}}
\newcommand{\eaa}{\end{eqnarray}}

\newcommand{\ba}{\bs\begin{eqnarray}}
\newcommand{\ea}{\end{eqnarray}\es}

\newcommand{\bt}[1]{\bs\label{#1}\begin{eqnarray}}
\newcommand{\et}{\end{eqnarray}\es}

\newcommand{\figlab}[2]{\begin{figure}\caption{#2}\label{#1}\end{figure}}

\newcommand{\paper}[6]{#1 , #2 #3 {\bf #4} , #5 (19#6)}

\newcommand{\refsec}[1]{Sec.~\ref{#1}}

\newcommand{\reffig}[1]{Fig.~\ref{#1}}

\newlength{\www}

\begin{document}
\draft
\preprint{ULDF-TH-1/3/00}

\title{Wave Functions for 
SU(2) \\ Hamiltonian Lattice Gauge Theory}

\author{Matteo Beccaria}

\address{
Dipartimento di Fisica dell'Universit\`a di Lecce, I-73100, Italy,\\
Istituto Nazionale di Fisica Nucleare, Sezione di Lecce}

\maketitle

\begin{abstract}
We study four dimensional SU(2) lattice gauge theory
in the Hamiltonian formalism by Green's Function Monte Carlo methods.
A trial ground state wave function is introduced to improve
the configuration sampling and we
discuss the interplay between its complexity and the 
simulation systematic errors. As a case study, we compare the
leading strong coupling approximation and an improved 4 parameters
wave function with $1\times 1$ and $1\times 2$ plaquette terms. Our
numerical results favors the second option.
\end{abstract}


\section{Introduction}
\label{sec:intro}

In the study of Lattice Gauge Theory, a major alternative to the 
usual Lagrangian formulation
is the Hamiltonian version of Kogut-Susskind~\cite{Hamiltonian}. 
This approach is relevant both for theoretical
reasons, e.g. universality checks, and for practical 
purposes because it seems the natural choice 
for static problems, like spectroscopy~\cite{Morningstar}.

An important feature of the Hamiltonian formulation is that it can
exploit analytical approximations of the ground state 
wave function to improve the numerical 
evaluation of physical quantities~\cite{Long}.
This well known procedure is called Importance
Sampling~\cite{ImportanceSampling}. However,
even if an accurate wave function is certainly welcome, 
on the other hand 
it can also require expensive calculations.
A careful analysis is therefore necessary to understand the interplay
between accuracy and performance and to determine 
which is the optimal  
complexity of the trial wave function.

A recent investigation of Hamiltonian methods in the 
numerical study of non Abelian gauge theories can be found in~\cite{Hamer96}
where the SU(2) model is studied in $2+1$ dimensions.
Here, we focus mainly on the  wave function issue and 
study the $3+1$ dimensional
SU(2) pure gauge lattice theory by  
a particular Green's Function Monte Carlo (GFMC)~\cite{Linden}
 algorithm 
that computes statistical
averages over an ensemble with a fixed number of random 
walkers~\cite{SR}. 
This algorithm has the desirable feature of allowing a particularly 
simple analysis of the systematic errors.

On general grounds and motivated by strong coupling
expansions~\cite{Guo94} we 
consider trial wave functions which are expressed by combinations
of gauge invariant Wilson loops with unknown coefficients $\boa$.
Two main issues are addressed: the cost of computing large loops
and the practical simultaneous
optimization of many parameters $\boa$.
About the latter, we test a recently proposed algorithm for the adaptive 
optimization of $\boa$. It is a method that has been successfully 
applied to the study of a very simple system with U(1) gauge 
symmetry~\cite{Mio2}
and that deserves a more detailed analysis on a realistic
test-bed like the non Abelian SU(2) four dimensional model.

As we shall see, the most serious systematic error is related to 
the finite population size, namely the number of walkers.
The extrapolation to an infinite population is difficult and 
much easier when an improved wave function is used.
Working with a four parameter trial function,
we present results for the extrapolated quantities and study the 
scaling of the string tension comparing our results with 
similar Lagrangian calculations.

The plan of the paper is the following:
in Sects.~\ref{sec:Model}, \ref{sec:SR} and \ref{sec:adaptive},
we present the GFMC algorithm in its general form 
for SU(N) lattice gauge theories with Importance Sampling, 
fixed number of random walkers and 
adaptive optimization of the trial wave function.
In~\refsec{sec:Results}, we apply the proposed algorithm to the SU(2)
model in four dimensions and discuss the numerical results.

\section{Review of GFMC for SU(N) gauge theory}
\label{sec:Model}

In this Section, we recall the Feynman-Kac-Nelson formula for the matrix
elements of the evolution operator associated with the SU(N)
Kogut-Susskind Hamiltonian. This formula provides probabilistic
representations of several interesting quantities. The GFMC algorithm is 
just one 
of its numerical implementations.

Let $U=\{U_l\}$ be the set of group elements, one for each 
link $l$; let $E=\{E_l^\alpha\}$ be the associated electric field 
defined by 
\be
[E^\alpha_l, U_{l'}] = T^\alpha U_l \ \delta_{l,l'},
\ee
where $\{T^\alpha\}$, $\alpha=1,\dots, N^2-1$ are SU(N)
generators. Finally, let $V(U)$ be a gauge invariant real
potential. 
The Kogut-Susskind Hamiltonian for SU(N) lattice gauge theory is
then
$H_{KS} = g^2 H$ where $g$ is the gauge coupling and
\be
H=H_0+V(U),\qquad H_0=\frac{1}{2} E^2\equiv
\frac{1}{2}\sum_{\alpha=1}^{N^2-1}\sum_l (E_l^\alpha)^2 .
\ee
For the Wilson action, $V(U)$ is 
proportional to the sum of the smallest $1\times 1$
plaquette loops $U_p$:
\be
V(U)=-\frac{1}{g^4}\sum_p\mbox{Tr} (U_p+U_p^\dagger),
\ee
but the present discussion holds for a general $V(U)$.

The matrix elements of the Euclidean evolution operator $\exp(-t H)$
in the basis of $U_l$ eigenstates admit the small $t$ expansion
\be
D(U'', U', \varepsilon) = \langle U'' | e^{-\varepsilon H} | U' \rangle = 
D_0(U'', U', \varepsilon)  
e^{-\varepsilon V(U')} + {\cal O}(\varepsilon^2) ,
\ee
where the factor
\be
D_0(U'', U', \varepsilon) = \langle U'' | e^{-\varepsilon H_0} | U'
\rangle ,
\ee
can be interpreted as a probability density for the random
transition $U'\to U''$ due to the relations: 
\be
D_0(U'', U', \varepsilon)>0\qquad \mbox{and} \qquad
\int D_0(U'', U', \varepsilon) dU'' = 1 ,
\ee
($dU$ is the SU(N) Haar measure).
Inserting intermediate states between
$U''$ and $U'$, we obtain as usual
\be
\label{poly}
D(U'', U', t) = \int \langle U'' | e^{-(t/(N+1)) H} | U_N \rangle
\cdots \langle U_1 | e^{-(t/(N+1)) H} | U' \rangle dU_1 \cdots
dU_N ,
\ee
and taking the $N\to\infty$ limit we recover the Feynman-Kac-Nelson
path integral representation of $D(U'', U', t)$:
\be
D(U'', U', t) = \int_{U(0)=U',\ U(t)=U''} {\cal D} U(t)
e^{-\int_0^t d\tau V(U(\tau))} ,
\ee
where the measure ${\cal D}U(t)$ stands for the
$N\to\infty$ limit of the average Eq.~(\ref{poly}) over random
group  {\em polygonals}
$(U_0=U', U_1, \dots, U_N,
U_{N+1}=U'')$.

The potential $V(U)$ fluctuates along the path $U(t)$ and determines the
error on the estimates of physical quantities. To see how this
happens, let us consider in this framework the calculation 
of the ground state energy $E_0$. 
The simplest procedure is to compute the limit
\be
E_0=\lim_{t\to +\infty} E(t, U'),
\ee
where
\be
E(t, U') = -\frac{d}{dt} \log \int D(U'', U', t) dU'',\qquad
\qquad U'\ \mbox{arbitrary}.
\ee
It is easy to check that
\be
E(t, U') = \frac{\int_{U(0)=U'} {\cal D}U(t)  e^{-\int_0^t d\tau
V(U(\tau))} V(U(t)) }{\int_{U(0)=U'} {\cal D}U(t) e^{-\int_0^t d\tau
V(U(\tau))}} \equiv \langle V(U(t)) \rangle ,
\ee
where $\langle\cdot\rangle$ is the average over {\em free} trajectories
on the group manifold weighted by the exponential term.
Hence, the fluctuations of $V$ control 
the noise in any estimator of $E_0$. Other problems
in the actual numerical evaluation of $\langle V\rangle$
can be discussed separately and will be considered
in the next Section.

The introduction of a trial wave function is a clever way to reduce
the fluctuations of $V$. To this aim, one considers a positive function
$\Psi_0(U) = \exp F(U)$ and the unitarily equivalent Hamiltonian
\be
\widetilde H=\Psi_0 [ H_0  + V(U) ] \Psi_0^{-1} ,
\ee
which turns out to be 
\ba
\widetilde H &=& \widetilde{H_0} + \widetilde V , \\
\widetilde{H_0} &=& \frac 1 2 E^2 + i E\cdot \nabla F, \\
E\cdot\nabla F &\equiv& \sum_{l, \alpha} E^\alpha_l \nabla_{\alpha, l} F \\
\widetilde V &=& V-\frac 1 2 \sum_{\alpha, l} (\nabla_{\alpha,l} F)^2
-\frac 1 2 \sum_{\alpha, l} \nabla_{\alpha, l}^2 F ,
\ea
where $\nabla_{\alpha, l}$ is the group invariant derivative on SU(N)
associated with link $l$.
The following approximate factorization of the $\widetilde H$ propagator holds
\be
\widetilde D = \widetilde D_0\ e^{-\varepsilon \widetilde V} + {\cal
O}(\varepsilon^2) , 
\ee
where $\widetilde D_0$ is the kernel of $\exp(-\varepsilon \widetilde{H_0})$
and satisfies the fundamental relations
\be
\widetilde D_0(U'', U', \varepsilon)>0\qquad \mbox{and} \qquad
\int \widetilde D_0(U'', U', \varepsilon) dU'' = 1 ,
\ee
(the important point is that $\widetilde{H_0}$ is 
{\em normal ordered} with all the $E$ operators acting from the left). 
As in the $F\equiv 0$ case, the finite time propagator $\widetilde D$
may be expressed in terms of a weighted average
\be
\widetilde D(U'', U', t) = \int_{U(0)=U',\ U(t)=U''}
\widetilde{\cal D} U(t)\ 
e^{-\int_0^t d\tau \widetilde V(U(\tau))} ,
\ee
where $\widetilde{\cal D}U$ is the measure determined by
the form of $\widetilde D$ at infinitesimal times.
The explicit first-order \underline{discrete}
 construction of the paths is done by choosing a 
time step $\varepsilon$ and updating 
$U_n \to U_{n+1}$ according to the rule
\be
\label{rwevol}
U_{n+1}=U_R(\varepsilon) \cdot U_D(\varepsilon, U_n) \cdot U_{n} ,
\ee
where $U_R(\varepsilon)
$ is a random SU(N) element distributed according to the 
heat kernel~\cite{Menotti,Chin} at time $\varepsilon$ and 
$U_D(\varepsilon, U)$ is the drift
\be
U_D(\varepsilon, U) = \exp(i \varepsilon T\cdot \nabla F(U)).
\ee

When $\Psi_0$ happens to be an exact
eigenstate of $H$ with eigenvalue $E$, we have $\widetilde
V\equiv E$ and the formula 
\be
E(t) = \langle \tilde V(U(t))\rangle = E,
\ee
provides the correct eigenvalue with zero variance, namely no statistical
error.
In other words, if we require $\widetilde V$ to be a constant, then
the problem is completely equivalent to solve the 
Schr\"odinger equation for the Kogut-Susskind Hamiltonian. In the following,
we shall try to obtain a constant $\widetilde V$ 
at least in some approximate
sense.

In the above discussion we focused on the ground state energy which is 
the simplest observable. A more general case is that of 
ground state matrix elements of diagonal operators in the basis 
of the link variables $U$:
\be
\langle \Omega\rangle = \langle 0 | \Omega(U) | 0 \rangle ,
\ee
where $|0\rangle$ denotes the ground state of $\widetilde H$.
They can be evaluated by computing
\be
\langle \Omega\rangle = \lim_{t_2\to +\infty}
\lim_{t_1\to +\infty} \Omega(t_2, t_1, U'),
\ee
where
\be
\Omega(t_2, t_1, U') =  \frac{\int_{U(0)=U'} {\cal D}U(t)  e^{-\int_0^{t_1+t_2}
d\tau
V(U(\tau))} V(U(t_1)) }{\int_{U(0)=U'} {\cal D}U(t) e^{-\int_0^{t_1+t_2} d\tau
V(U(\tau))}}.
\ee
A similar formula can be also devised for many time correlation functions
of operators in the Heisenberg representation.

\section{Stochastic Reconfiguration}
\label{sec:SR}

In this Section, we discuss the actual calculation of the above
mentioned averages over trajectories. Here, the
trial wave function $\Psi_0$ will not play any role.

A naive implementation of the Feynman-Kac-Nelson representation
generates group trajectories according to the measure ${\cal D}U$
and weights them with a potential dependent weight. Such
weight increases or vanishes exponentially fast with respect to the
trajectory length $t$. This is a serious numerical problem referred to
as the ``exploding variance problem''. The name stems from the fact
that the weight variance of a collection of random walkers explodes as
$t\to +\infty$ overwhelming the numerical capabilities of any 
machine~\cite{Hetherington}.

A possible solution to this problem consists in killing and cloning
the random walkers with definite rules in
order to delete the walkers with low weights and duplicate the others.
The main disadvantage is that the size of the walker
population varies in time and may become unstable.

The recently proposed Stochastic Reconfiguration Algorithm~\cite{SR}
is a simple implementation of the killing and branching idea, but with
the desirable feature of dealing with a population with a fixed
number of walkers. Its coding is therefore much simpler.

To this aim, a finite collection of $K$ walkers, an ensemble,  is introduced
\be
\ens = \{(U^{(n)}(t), \omega^{(n)}(t))\}_{1\le n\le K} ,
\ee
where the weights $\omega^{(n)}$ are defined by 
\be
\omega^{(n)}(t)=\exp -\int_0^t d\tau V(U^{(n)}(\tau)) .
\ee
The variance of the weights over the ensemble $\ens$ is 
\be
W(t)=\mbox{Var}\ \omega(t) = \frac{1}{K}\sum_{k=1}^K (\omega^{(k)}(t))^2 -
\left(\frac{1}{K}\sum_{k=1}^K \omega^{(k)}(t)\right)^2 .
\ee
The average of a function $f(U)$ over $\ens$ is defined as 
\be
\langle f\rangle_\ens = \frac 1 K \sum_{k=1}^K f(U^{(k)})
\omega^{(k)} .
\ee
When $W(t)$ becomes too large, our aim is to transform $\ens$ in a
new ensemble $\ens'$ with a rule $\ens\to\ens'$ such that 
the variance $W$ is reduced while the averages are kept constant. 
Actually, this is a formidable task, but we can achieve it
at least approximately, namely in the $K\to\infty$ limit. 
To this aim, we extract
$K$ independent integers $n_1, \dots, n_K$ with $n_i\in\{1, \dots, K\}$
and probabilities
\be
\mbox{Prob}(n_i=k) = \frac{\omega^{(k)}}{\sum_{k'=1}^K \omega^{(k')}} .
\ee
We then define $\ens'$ by 
\be
\ens' = \{(U^{(n_i)}(t), 1)\}_{1\le i\le K} .
\ee 
In other words, we replace the information carried by the weights with that 
encoded in the multiplicities of unit weight walkers.
By the large numbers law we expect
\be
\langle f \rangle_\ens - \langle f \rangle_{\ens'}\stackrel{K\to\infty}{\longrightarrow} 0
\ee
In actual calculations, there is therefore a systematic error which is
removed by performing simulations with different values of $K$ and
somehow extrapolating the  $K\to \infty$ limit.

The calculation of observables when stochastic reconfiguration is in progress
does not present any additional problem. The only point we stress is that, 
for a general operator $\Omega$, the weights at time 
$t_1+t_2$ are used to weight $\Omega(t_1)$ and therefore
one must keep track for each reconfigured walker of its states 
at that past time (usually, a small number of updates is enough).

\section{Adaptive Optimization of the Trial Wave Function}
\label{sec:adaptive}

Let us consider a trial wave function $\Psi_0(U, \boa)$ depending on 
some parameters
$\boa~=~(a_1, \dots, a_p)$ and the GFMC calculation of
an observable quantity that, for simplicity, we choose to be 
the ground state energy $E_0$. 

We do not discuss the systematic error associated with 
$\varepsilon$ appearing in Eq.~(\ref{rwevol}).
The choice of $\varepsilon$ is related to the
approximation involved in the operator splitting
 and also in the evaluation of the SU(N) heat
kernel. A common model independent choice is to take $\varepsilon$
enough small to approximate the diffusion over SU(N) with that on
the flat tangent space at the identity. Problems may then occur only if
$\nabla F$ becomes too large, a constraint that can be checked during
the simulation. In the following we shall choose a
reasonably small $\varepsilon$ and work under the hypothesis that the
$\varepsilon\to 0$ extrapolation does not change our conclusions on
the trial wave function dependence.

A simulation with a population of $K$ walkers will provide after $S$
Monte Carlo steps an approximate estimator $\hat E_0(S, K, \boa)$ of $E_0$.
This is a random variable with the property
\be
\langle \hat E_0(S, K, \boa) \rangle = E_0 + c_1(K, \boa),
\ee
and
\be
\mbox{Var}\ \hat E_0(S, K, \boa) = \frac{c_2(K, \boa)}{\sqrt{S}},
\ee
where $\langle\cdot\rangle$ denote the average over the Monte Carlo
realizations. The finite population functions $c_i(K, \boa)$ 
satisfy
\be
\lim_{K\to\infty} c_i(K, \boa) = 0 ,
\ee
where, by a self-averaging argument, it is reasonable to 
guess the asymptotic form $c_2(K, \boa)\sim \gamma(\boa)/\sqrt{K}$
for large $K$.
The average of $\hat E_0$ extrapolated at $K\to\infty$ 
is exact and independent on
the trial parameters $\boa$. On the other hand, the function
$c_2(K,\boa)$ may be strongly dependent on them. 
Since $c_2$
determines the statistical error at fixed computational time, it is
important to minimize it.

If the family of functions $\Psi_0(U, \boa)$ 
includes the exact ground state at the special point $\boa=\boa^*$ then
we know that 
\be
c_1(K, \boa^*)=c_2(K, \boa^*)=0 .
\ee
In a less optimal situation one has the two possibilities of
minimizing $c_1$ or $c_2$. Motivated by~\cite{Umrigar88}, we
choose to pursue the second alternative and seek a minimum of
$c_2$. 

Moreover, we propose to optimize $\boa$ adaptively.
This can be done in the following way: the
constant $c_2$ is related to the non zero fluctuations of 
$\widetilde V$ defined by 
\be
\widetilde V(U, \boa) = V(U)-\frac 1 2 (\nabla F(U,
\boa))^2-\frac 1 2 \nabla^2 F(U, \boa),
\ee
where $\Psi_0(U,\boa) = \exp F(U, \boa)$.
The ensemble variance of $\widetilde V$ is a function of $\boa$
\be
\mbox{Var}_\ens\ \widetilde V = \frac{1}{K}\sum_{k=1}^K \widetilde
V(U^{(k)},\boa)^2 \omega^{(k)}-\left(\frac{1}{K}\sum_{k=1}^K \widetilde
V(U^{(k)},\boa) \omega^{(k)}\right)^2 .
\ee
Our proposal is to update $\boa$ according to the simple equation
\be
\label{boaevol}
\boa_{n+1} = \boa_n -\eta \nabla_\boa \mbox{Var}_\ens\ \widetilde V ,
\ee
where $\eta$ is a constant parameter and $\boa_n$ is the value of
$\boa$ at the n-th update.
In other words, following the spirit of~\cite{Umrigar88}, we implement
a local minimization of the weight variance as a driving mechanism for
the free parameters.
It is important to remark that it is incorrect to look for 
the special parameters $\boa$ that minimize
$\mbox{Var}_\ens\ \widetilde V$ for a \underline{fixed} ensemble $\ens$. 
In fact, the equilibrium 
distribution of the link variables in the ensemble do depend
on $\boa$ and the addition of Eq.~(\ref{boaevol}) to the Monte Carlo
core establishes a non trivial feedback between 
the two sets of variables.

The coupled set of equations Eqs.~(\ref{rwevol},\ref{boaevol}) for the 
evolution of $\boa$ and the
random walkers is non linear and discrete. The above procedure is
reasonable, but a test in explicit examples is required to check the
convergence and stability of the method. A first analysis is described
in~\cite{Mio2} for a toy model with U(1) gauge symmetry in low
dimensions.
Here we consider the less trivial non Abelian four dimensional case.

\section{Simulation and Results}
\label{sec:Results}

\subsection{General Setup}

We simulate the SU(2) lattice gauge theory on 
a spatial $4^3$ lattice. 
The temporal extension is in principle infinite.
Of course, the meaning of this statement is that in the Hamiltonian formulation
we are projecting onto the exact ground state since we are applying 
many times an approximate form of $\exp(-\varepsilon H)$.
In all our simulations $\varepsilon = 0.01$.
Another parameter that we keep fixed for simplicity is the number of 
elementary time evolutions
between two stochastic reconfigurations. We choose 10 steps which is
roughly the decorrelation time of the lattice configuration.
A priori, a better choice of this parameter can improve the
convergence of the $K\to\infty$ extrapolation, but we checked that in
our range of parameters, the sensitivity on the choice of $K$ is quite
small. 
We also fix the learning parameter $\eta=0.001$ for the adaptive determination
of the wave function coefficients.

The parameters that remain free are therefore: the
number $K$ of walkers, the number of coefficients in the 
ground state wave function and of course the coupling constant $g$ 
which is needed to discuss the 
possible scaling of the measured quantities.
In more details, for each value $K=2$, $10$, $50$, $100$ and for each
choice of the trial wave function we run the algorithm at five
different coupling constants
\be
\label{range}
\lambda = 4/g^4 = 0.8,\ 0.9,\ 1.0,\ 1.1,\ 1.2 ,
\ee
and measure the ground state energy $E_0$ and all the
Wilson loops $W_{IJ}$ with area $IJ \le 9$. We recall that the
variable $4/g^4$ is natural in the strong coupling expansion analysis.

\subsection{Adaptive Optimization}

The first topic that we discuss is the adaptive optimization of the
free parameters $\boa$ in the trial wave function.
Motivated by the long
wavelength strong coupling analysis of~\cite{Guo94}, 
we consider the following form
of $F$
\be
F = c_1 W_{1\times 1} + c_2 W_{(1\times 1)^2} + c_3 W_{1\times 2, \rm planar}
+c_4 W_{1\times 2, \rm bent} ,
\ee
where $W_{1\times 1}$, $W_{(1\times 1)^2}$, $W_{1\times 2, \rm planar}$
and $W_{1\times 2, \rm bent}$ are the sum of all the plaquettes of the kind 
shown in~\reffig{fig:1}. The coefficients $c_1$, $c_2$, $c_3$ and $c_4$ may be 
approximated as in~\cite{Guo94}
 with a power series in $1/g^2$ computed on the infinite
lattice. Here, we exploit the dynamical system~Eq.(\ref{boaevol})
to determine
the finite size optimal $c_i$. In the analysis we discuss three
specific scenarios that we characterize by the level $l=0$, $1$ or
$4$,
the number of free parameters:
\begin{enumerate}
\item[] $l=0$:  $F\equiv0$, no Importance Sampling;
\item[] $l=1$:  $c_2=c_3=c_4=0$, one free parameter;
\item[] $l=4$:  all $c_i$ are optimized.
\end{enumerate}

In~\reffig{fig:7}, we show for example 
the evolution of $c_1$ with Monte Carlo time
in the case $K=2$, $1/g^2=0.5$ and $l=1$. The coefficients $c_i$
typically stabilize on values adaptively computed by the algorithm 
with small fluctuations that can be averaged in the long time limit.
The initial values of $c_i$ for $K=10$, $50$, $100$ are taken from the 
run with the nearest smaller $K$ in order to reduce thermalization
times. The procedure turns out to be stable at each considered
$K$ and $g$. The fluctuations are bigger at smaller $g$, that is in the 
critical limit, while the equilibrium values 
agree reasonably well with the strong coupling approximation
at large $g$. If the fluctuations were too noisy, then Stochastic Gradient 
Approximation techniques~\cite{Harju97} could 
be applied by gradually reducing $\eta$ in Eq.~(\ref{boaevol}).

\subsection{$K\to\infty$ extrapolation}

For each level $l$, we
consider simulations with 
$K=2$, $10$, $50$ and $100$. The $K\to\infty$ limit
does not depend on $l$, but the rapidity of the convergence yes.
In Figs.~(\ref{fig:2}-\ref{fig:3}) 
we plot the $K$ dependent estimates of $E_0$ and
 $W_{1\times 1}$ for $l=0$, $1$, $4$ at $1/g^2=0.5$. In the left plots
we see clearly that for $l=0$ and $l=1$ the extrapolation $K\to \infty$ is
largely uncertain. Unless we know the asymptotic behavior in $K$, we cannot
safely determine the correct limit with the present data. 
A non rigorous guess is a 
power law behavior with finite $K$ corrections of the form
\be
c_1(K, \boa) = \frac{c_1(\boa)}{K^{\alpha(\boa)}} +
o(K^{-\alpha}), \qquad \alpha(\boa)>0,
\ee
that has been observed empirically~\cite{Mio2}.
Such behavior can be matched to the $K=10$, $50$, $100$ data,
but its choice is rather arbitrary and, to be supported, 
would require more points at larger $K$.
On the other hand, data computed in the $l=4$ scheme show
a plateau beginning at $K\simeq 50$ and data at $K=100$ will therefore be 
used as an estimate of the $K\to\infty$ limit.

To check consistency of the runs at different $K$ and to evaluate 
quantitatively the improvement gained with larger $K$, we 
introduce an effective $K$ 
\be
K'= K/\lambda(l),
\ee
depending on the rescaling factor $\lambda(l)$.
In the right plots of Figs.~(\ref{fig:2}-\ref{fig:3}), 
we show how the curves for $E_0$ and $W_{1\times 1}$ measured
at different $K$ collapse on a single curve when shown as
functions of $K'$. The values of $\lambda(l)$ are 
\be
\lambda(0) = 1.6\cdot 10^3, \qquad \lambda(2) = 5,\qquad
\lambda(4)\equiv 1 .
\ee
The raw simulation, $l=0$, is very poor, as expected.
On the other hand, 
the one parameter wave function gives the same accuracy of the $l=4$ one
when the number of random walkers is roughly 5 times bigger.

We now determine the computational cost as a function of $l$.
Let $n$ be the number of Monte Carlo iterations; the statistical
error in the measure of a given observable 
and the CPU time are respectively given by
\be
\label{errors}
\varepsilon_{stat} = \frac{\gamma_l}{\sqrt{nK}},\qquad
t_{CPU} = \frac{nK}{v_l},
\ee
where $\gamma_l$ and $v_l$ are normalization constants expected to 
be a decreasing function of $l$. Of course, $\gamma_l$ depends on the chosen 
observable. From Eq.~(\ref{errors}) we deduce that 
\be
t_{CPU} = \frac{\gamma_l^2}{v_l}\frac 1 {\varepsilon_{stat}^2}
\ee
is independent on $K$. From our data, we can give the rough
estimate of the ratios $v_1/v_4$ and $\gamma_1/\gamma_4$ when the
coupling is $1/g^2\simeq 0.5$. We find
\be
v_1/v_4 \simeq 8 ,\qquad
\gamma_1/\gamma_4 \simeq 2 ,
\ee
and therefore $t_{CPU, l=4} \simeq\ 2\ t_{CPU, l=1}$. 
Of course one must also remember that, as discussed above, $l=1$
does not show a clear plateau when $K<100$ and that 
very large $K$ cannot be used in $3+1$ dimensions 
because of storage limitations.
Larger Wilson loops show a similar behavior with slightly better improvement
in the case $l=4$.

In this particular model and at the considered volume and couplings
we summarize the comparison between $l=4$ and $l=1$ at \underline{
fixed systematic and statistical} 
errors by saying that the improved wave functions are 
twice slower, but require a storage $\simeq 5$ times smaller.
In the following, we shall pursue the $l=4$ choice because only in that 
case, our data can be safely considered asymptotic with respect to the
$K\to\infty$ limit.

\subsection{Scaling of the string tension}

The $l=4$, $K=100$ measurements of $E_0$, the Wilson loops with 
area $\le 9$ and the coefficients $c_i$
are shown in Figs.~(\ref{fig:4}-\ref{fig:6}) as functions of $1/g^2$. 
The energy measurements agree well with the second order 
strong coupling expansion
\be
E_0 = \frac{4}{g^4}-\frac{8}{3 g^8} + {\cal O}\left(\frac{1}{g^{12}} .
\right)
\ee
An important check of the simulation requires an attempt to investigate 
scaling.  The easiest observable quantity with simple Renormalization
Group behavior is the string tension $\sigma$ that we
obtain from Creutz's ratios
\be
\sigma = \lim_{I,J\to\infty} \sigma_{I,J} , 
\qquad
a^2 \sigma_{I,J} = -\log\frac{W_{I,J}W_{I+1,J+1}}{W_{I,J+1}W_{I+1,J}} .
\ee
This must be compared with the asymptotic scaling prediction
\be
a(g) = \frac 1 {\Lambda_H} f(g),
\qquad f(g) = \left(\frac{24\pi^2}{11g^2}\right)^{51/121}
\exp\left(-\frac{12\pi^2}{11g^2}\right) .
\ee
Even if we are work on a lattice with rather small spatial
extension, it is interesting to compare our results with those
obtained in the Lagrangian formulation. We choose the measurements 
in~\cite{Rebbi} as representatives of what can be obtained at values of 
the coupling in the range Eq.~(\ref{range}).
In~\reffig{fig:8} we show $a\sqrt{\sigma_{1,2}}$ and $f(g)$. 
Loops larger than $2\times 3$ cannot be used since in our 
computation they have too large errors.

The Lagrangian measurement is 
$\Lambda_L = (1.19\pm 0.15)\cdot 10^{-2}\sqrt{\sigma}$.
Taking into account the SU(2) $\Lambda$ parameter ratio
$\Lambda_H/\Lambda_L = 0.84$~\cite{Hasen}, 
we obtain the band shown in the Figure. 
As one can see, the Hamiltonian measurement is slightly larger, but 
reasonably consistent with the Lagrangian one.



\section{Conclusions}
\label{sec:conclusions}

The aim of this paper has been a study of the non perturbative behavior
of the $SU(2)$ lattice gauge theory in $3+1$ dimensions by Hamiltonian
Monte Carlo methods.
Our results can be summarized into two main statements. 

First,
we have shown that the algorithm~\cite{Mio2} 
is able to adaptively optimize the 
many parameters ground state trial wave functions needed to guide 
GFMC simulations of realistic models.

Second, we have discussed to what extent the use of improved 
wave functions is actually necessary, at least with the considered volume and
couplings. 
From the purely computational
point of view our data show that wave function improvement 
reduces the algorithm performance, 
but permits a substantial gain in storage.
This means that we are allowed to work with a 
smaller number of walkers, $K$. 
This can be an important advantage, especially on larger lattices or 
at smaller couplings where the minimum $K$ at which a safe
extrapolation to the $K\to \infty$ limit is feasible is expected to 
increase.

As a non trivial check of the algorithm and of the strategy for the 
removal of the finite $K$ 
systematic error, we studied 
the string tension with a four parameter trial
wave function and found a scaling behavior compatible with 
similar Lagrangian measurements.

Our investigation has been performed on a cluster of small 
computers and we could not attempt to measure glueball mass ratios.
At this stage, the Hamiltonian formulation does not offer any
special advantage with respect to the Lagrangian one. 
The storage gain associated to the absence of 
the temporal dimension is compensated by the memory required to 
manage the walkers ensemble.
In practice, we did not examine finite size scaling of the algorithm, 
a task that we leave for future studies.
We stress however that many of the advanced techniques that are currently 
used in Lagrangian calculations~\cite{Morningstar},
can be applied in the Hamiltonian formalism as well. Examples are 
the use of refined glueball operators or the improvement 
programme that has been recently extended to 
Kogut-Susskind Hamiltonians~\cite{Luo99}.

\acknowledgements
I whish to thank Prof. G. Curci for many useful discussions on the 
numerical simulation of lattice field theories.

\figlab{fig:1}{Plaquettes appearing in the $l=4$ trial wave
function. A sum over the lattice is understood.}

\figlab{fig:7}{Monte Carlo history of the trial wave
function coefficient $c_1$ at $1/g^2 = 0.5$ and
$K=2$. The initial state of the system is cold.}

\figlab{fig:2}{Dependence of $E_0$ on $K$ with and without
reparametrization $K\to K'$ for the three schemes $l=0$, $l=1$ and
$l=4$. The coupling constant is $1/g^2=0.5$.
}

\figlab{fig:3}{As \reffig{fig:2}, but for the $1\times 1$ Wilson loop.}

\figlab{fig:4}{Coupling constant dependence of $E_0$ at
$l=4$, $K=100$.}

\figlab{fig:5}{Coupling constant dependence of the trial wave function
coefficients at $l=4$, $K=100$.}

\figlab{fig:6}{Coupling constant dependence of the Wilson loops at
$l=4$, $K=100$.}

\figlab{fig:8}{Scaling plot of the string tension. 
Comparison between $a\sqrt{\sigma_{1,2}}$ and the asymptotic 
Renormalization Group prediction. 
The band corresponds to the Lagrangian measurement as explained 
in the text}.


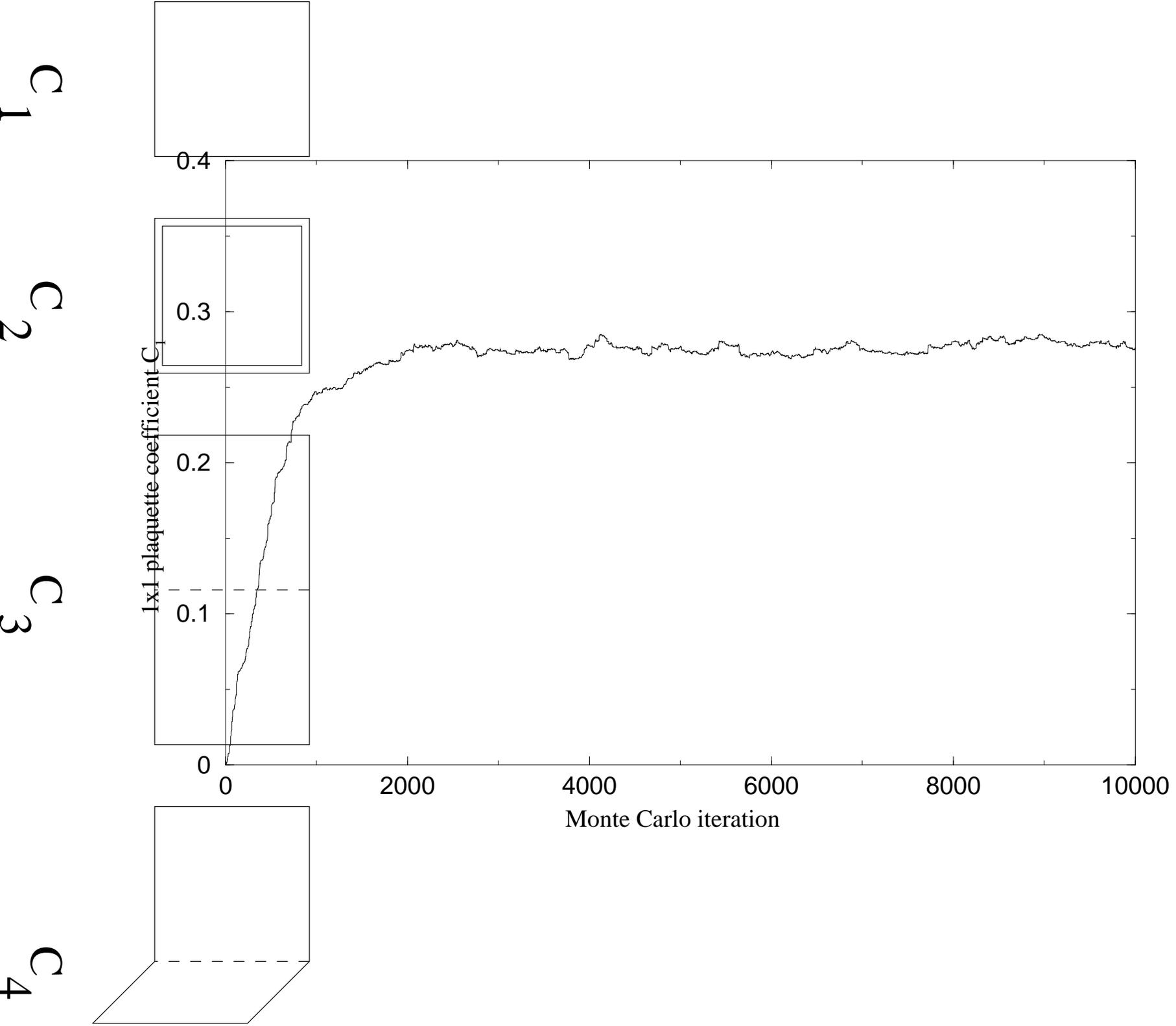
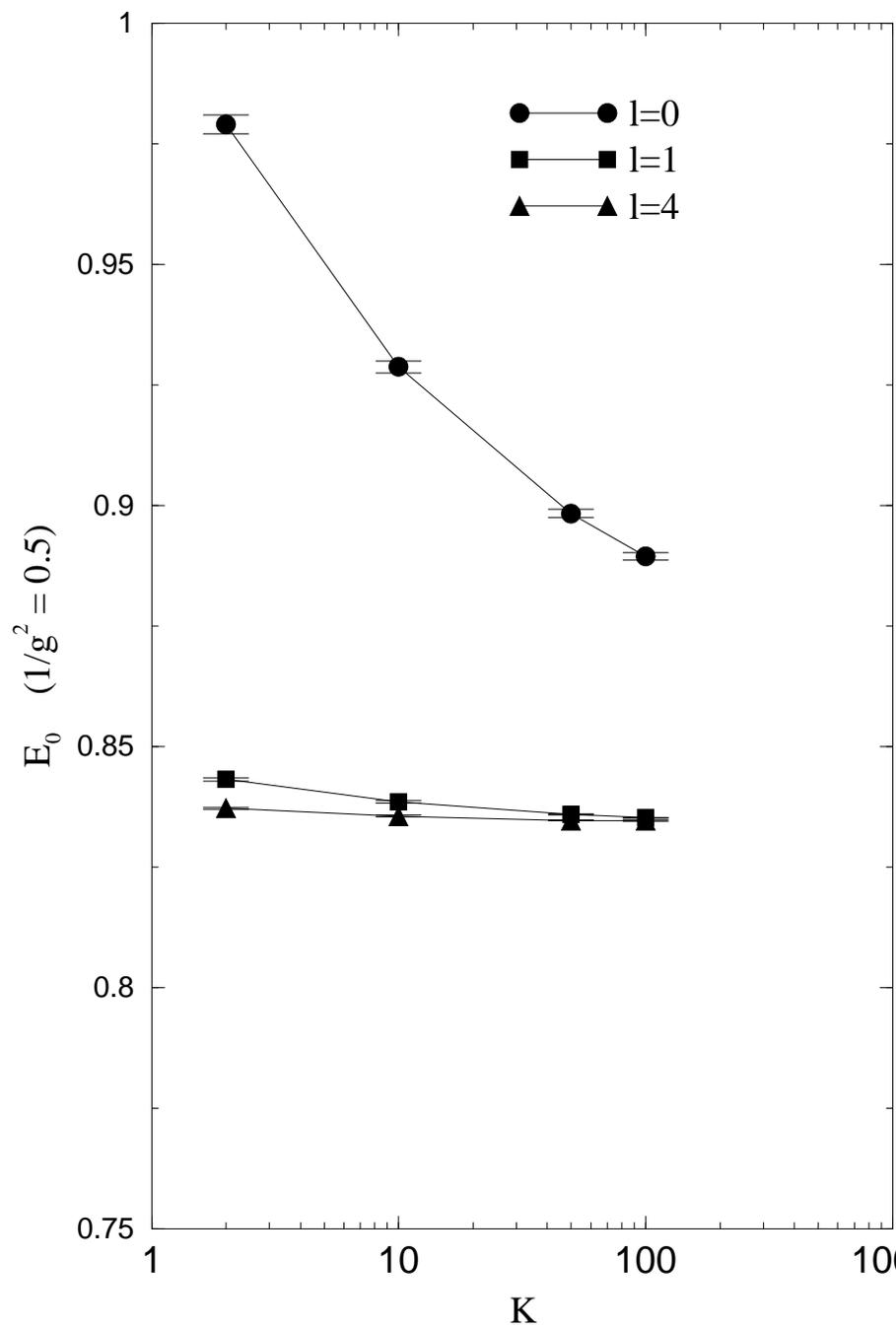
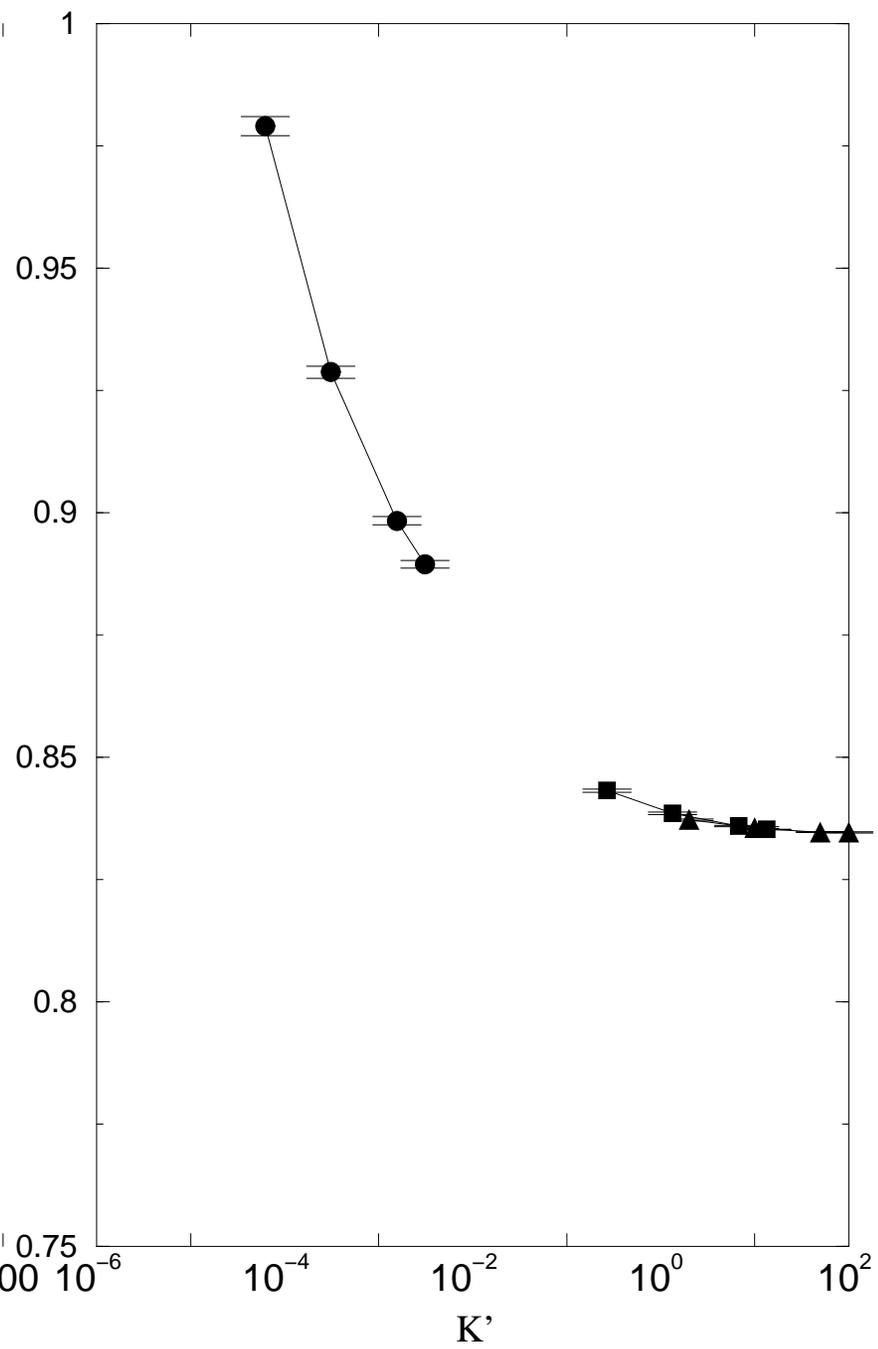
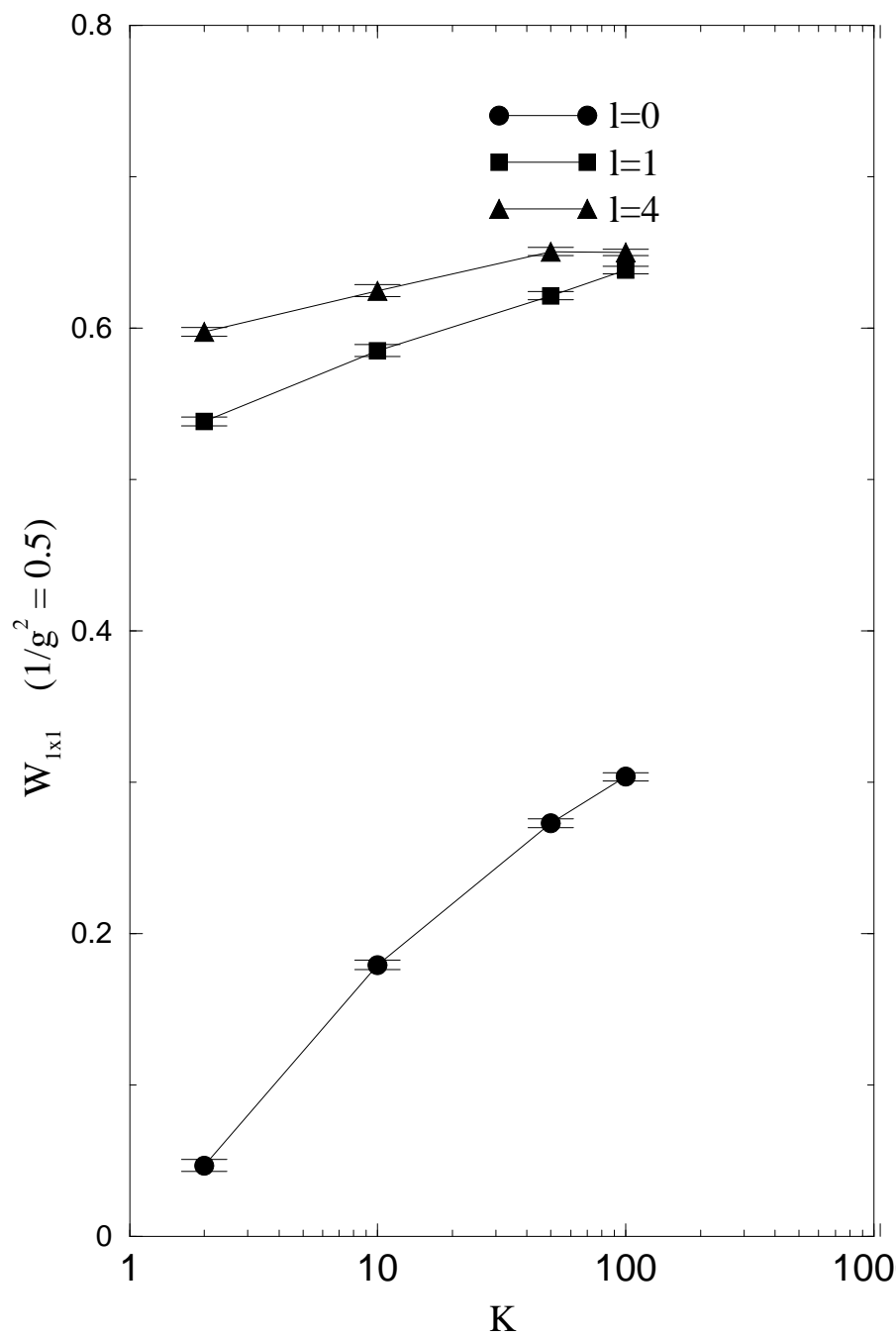
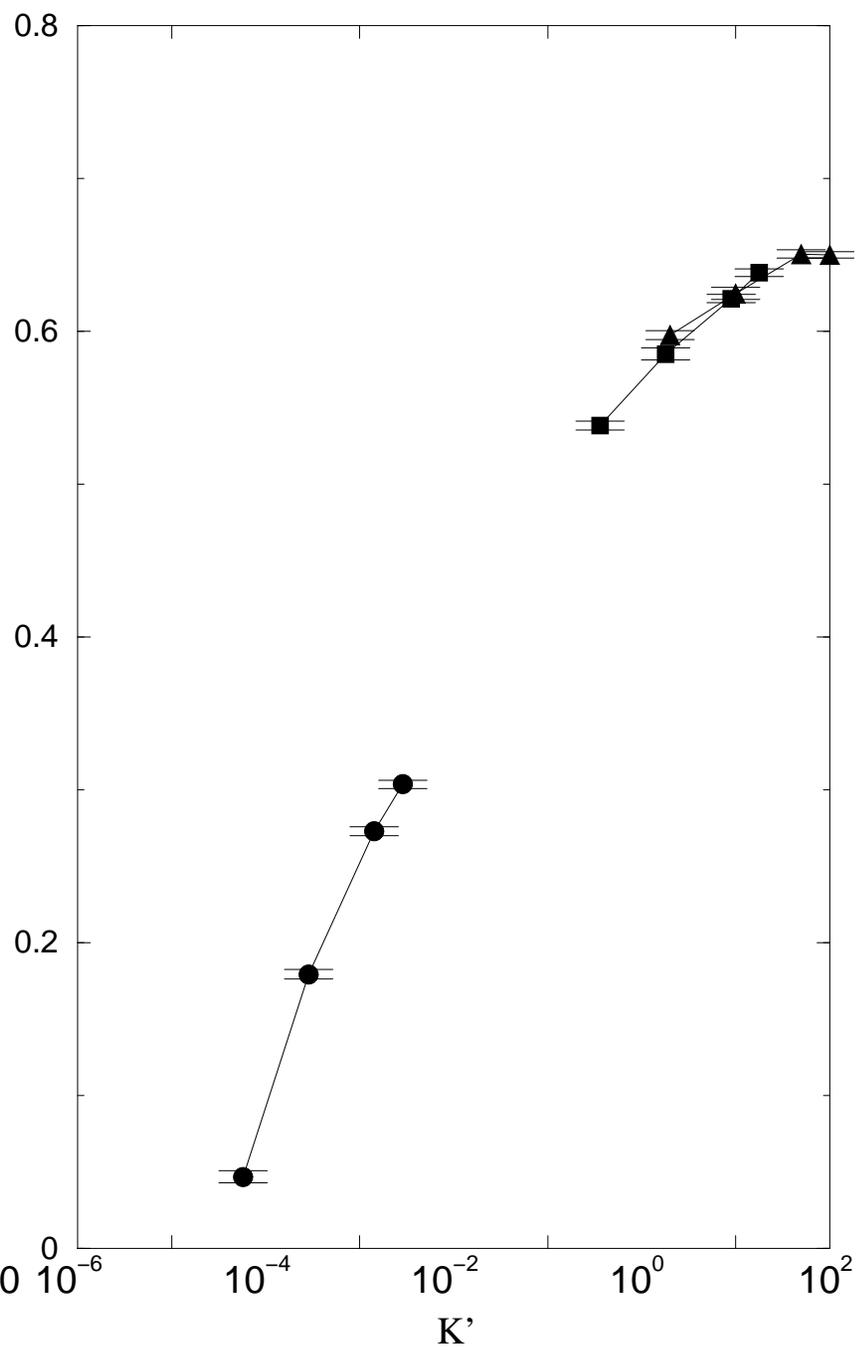
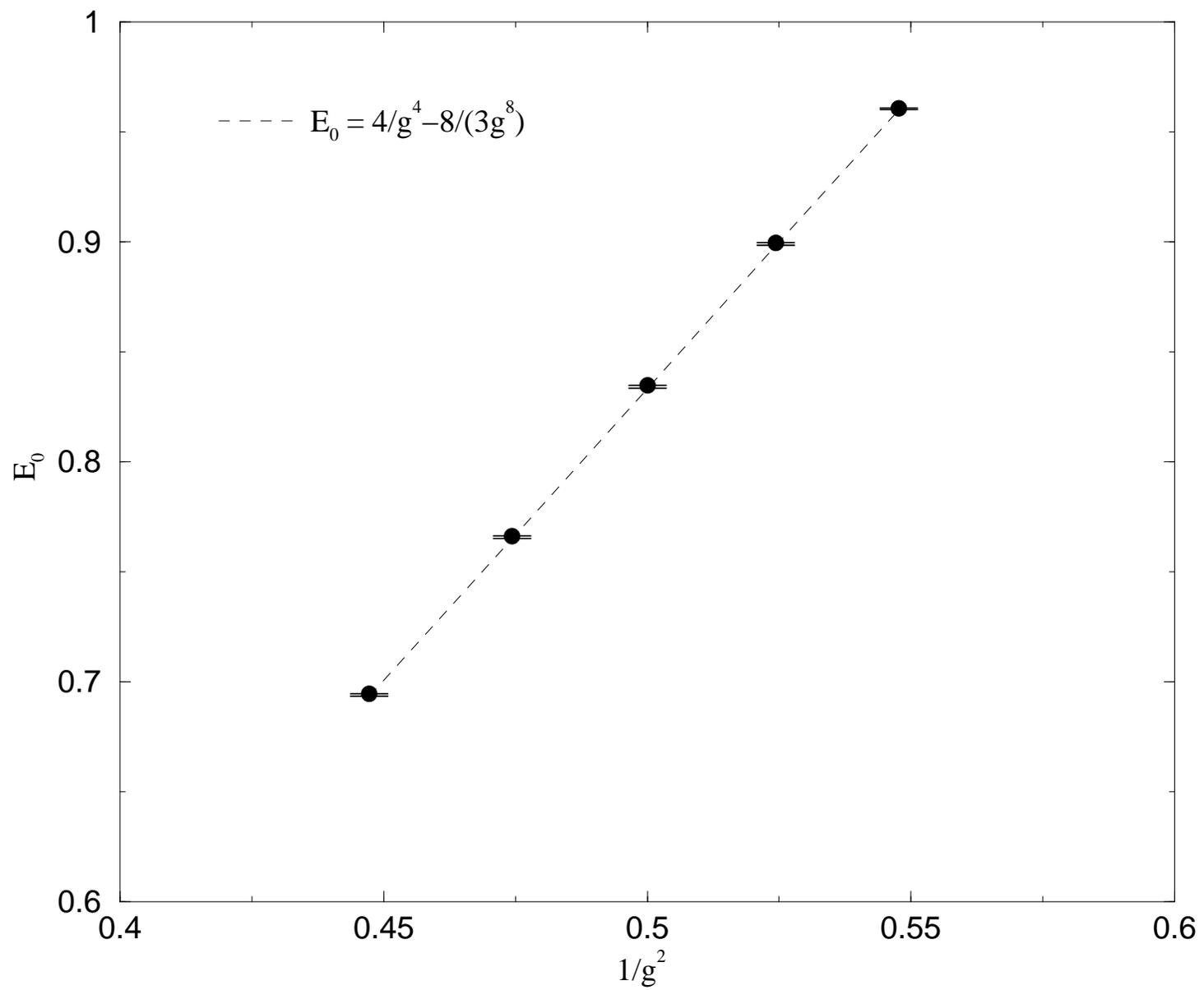
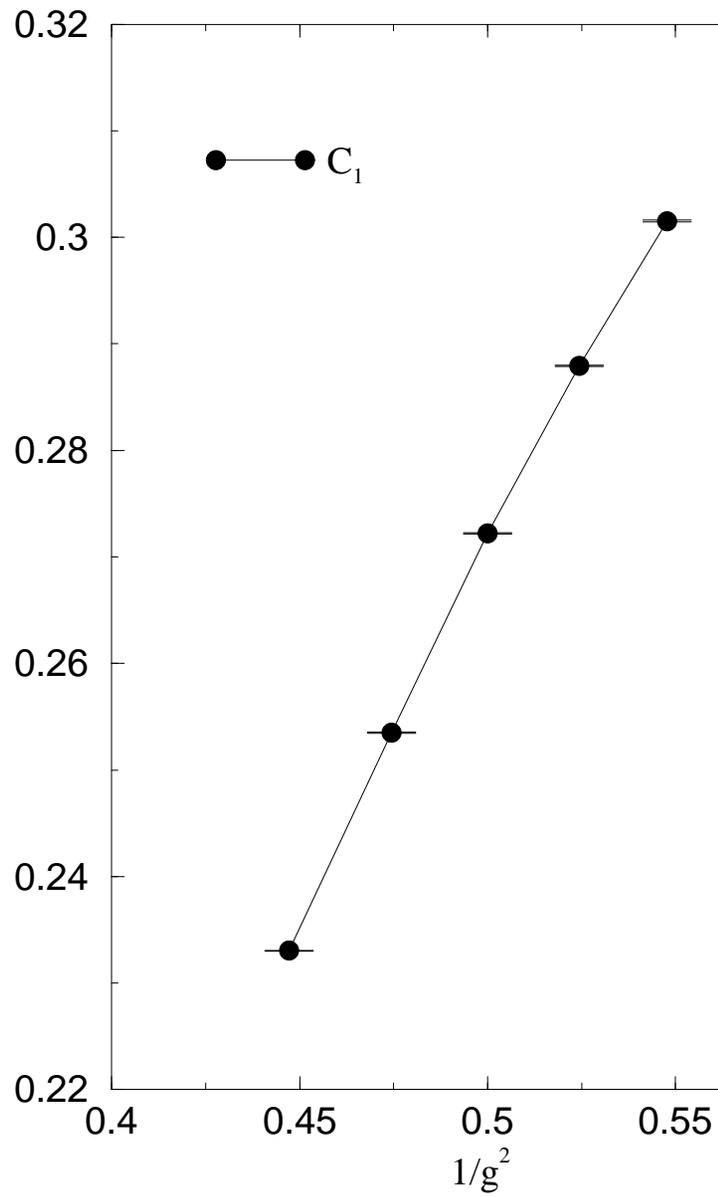
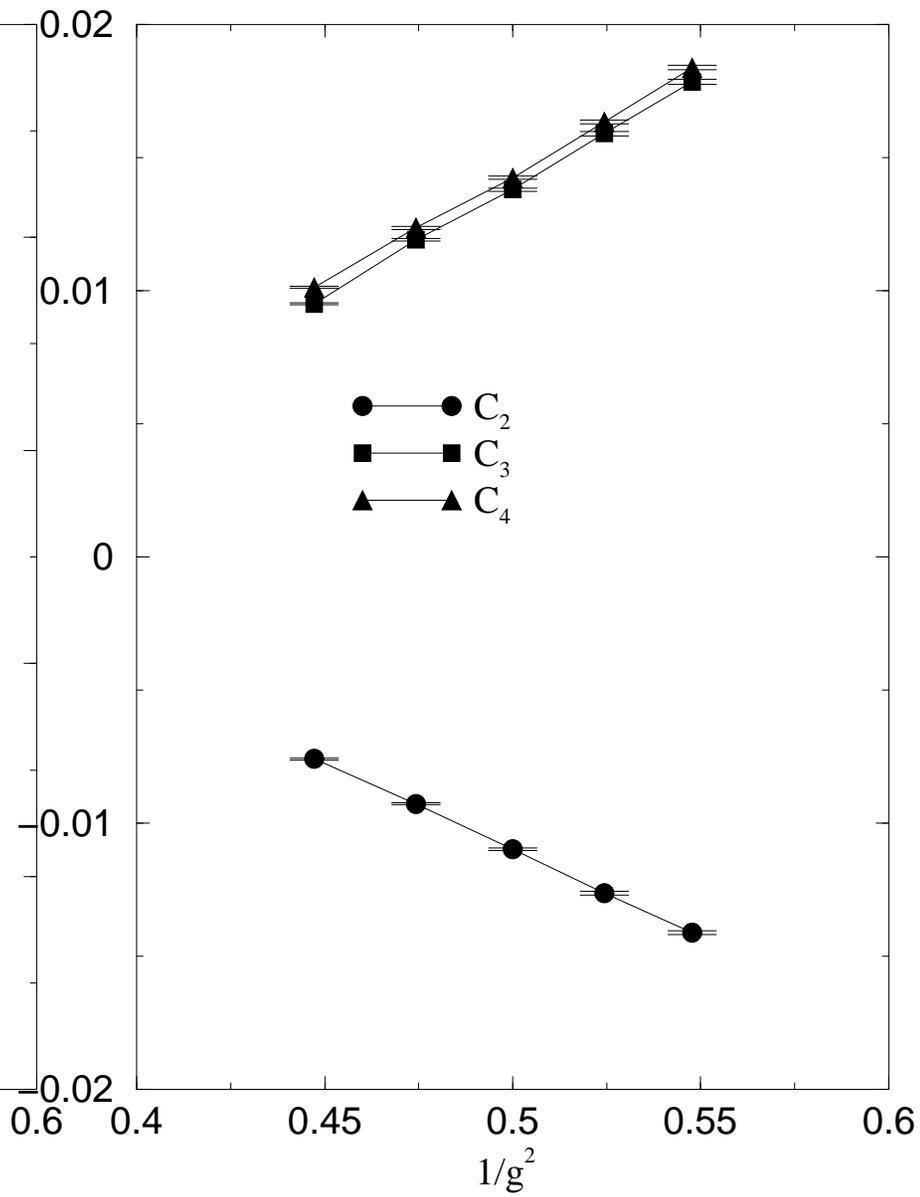
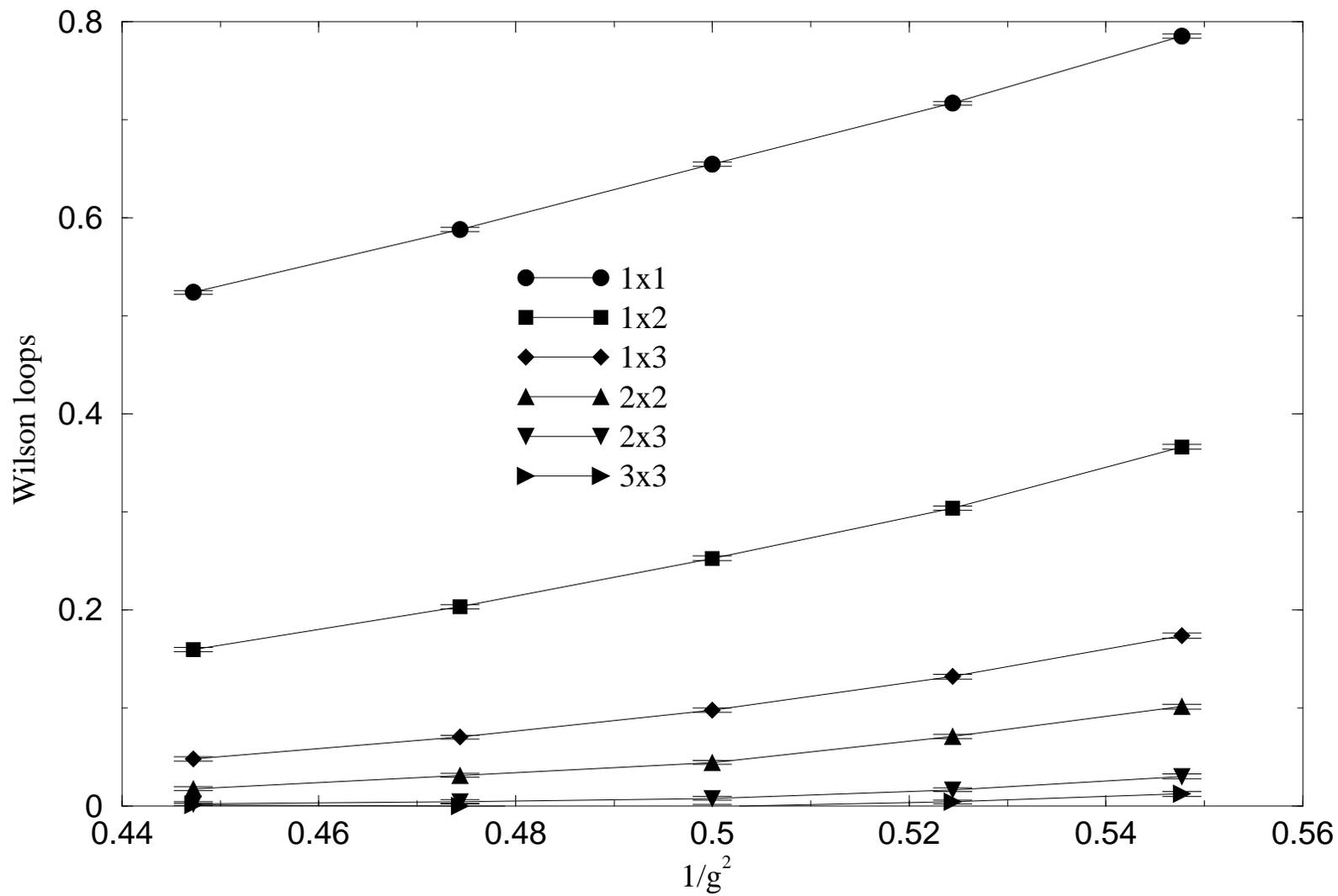
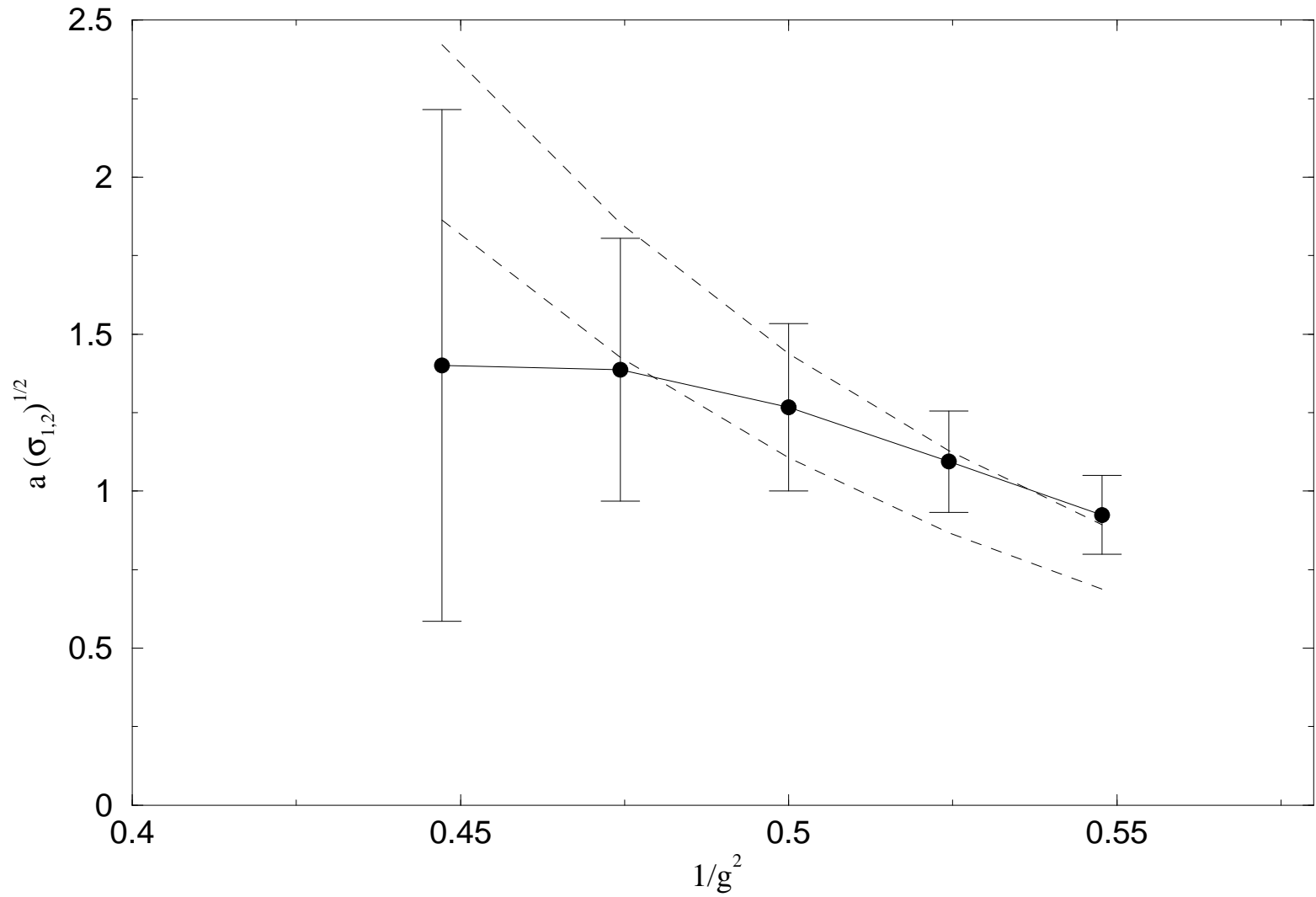
\end{document}